\documentstyle[aps,prb,epsfig,floats]{revtex}
\begin{document}
\draft
\wideabs{
\title{``Quasi 2-D'' Spin Distributions in II-VI Magnetic Semiconductor
 Heterostructures: Clustering and Dimensionality}
\author{S. A. Crooker}
\address{National High Magnetic Field Laboratory - Los Alamos National
Laboratory, NM 87545}
\author{N. Samarth}
\address{Department of Physics, Pennsylvania State University,
 University Park PA 16802}
\author{D. D. Awschalom}
\address{Department of Physics, University of California,
Santa Barbara CA 93106}
\date{Received 19 July 1999}
\maketitle
\begin{abstract}
Spin clustering in diluted magnetic semiconductors (DMS) arises
from antiferromagnetic exchange between neighboring magnetic
cations and is a strong function of reduced dimensionality.
Epitaxially-grown single monolayers and abrupt interfaces of DMS
are, however, never perfectly two-dimensional (2D) due to the
unavoidable inter-monolayer mixing of atoms during growth. Thus
the magnetization of DMS heterostructures, which is strongly
modified by spin clustering, is intermediate between that of 2D
and 3D spin distributions. We present an exact calculation of spin
clustering applicable to {\it arbitrary} distributions of magnetic
spins in the growth direction.  The results reveal a surprising
insensitivity of the magnetization to the form of the intermixing
profile, and identify important limits on the maximum possible
magnetization. High-field optical studies of heterostructures
containing ``quasi-2D'' spin distributions are compared with
calculation.
\end{abstract}
%Do not delete this line
} \narrowtext Spin clustering is ubiquitous in II-VI diluted
magnetic semiconductors (DMS), resulting in reduced effective
magnetizations at low magnetic fields and magnetization steps at
high fields.\cite{Shapira} Clear predictions can be made for the
number and type of spin clusters in 3D systems ({\it e.g.}, bulk
Cd${_{1-x}}$Mn${_x}$Se), where the distribution of magnetic
Mn${^{2+}}$ cations is random and isotropic. With the advent of
molecular beam epitaxy (MBE) and other techniques for
monolayer-by-monolayer growth of DMS heterostructures, spin
clustering in systems of reduced dimensionality has enjoyed much
recent interest. It is well
established\cite{Grieshaber,Gaj,Ossau,Lemaitre} that spin
clustering (arising mainly from an antiferromagnetic exchange
between neighboring magnetic cations) should be greatly reduced in
two-dimensional systems such as abrupt interfaces or discrete
monolayer planes, leading to enhanced paramagnetism. However,
experiments show that perfect 2D interfaces and monolayers are
never realized due to the inevitable inter-monolayer mixing of
atoms during MBE growth, which smears the magnetic cations over
several monolayers. Common mechanisms include segregation (mixing
between the monolayer being grown and the underlying monolayer)
which leads to roughly exponential magnetic profiles, and
diffusion (which can arise from, {\it e.g.}, high growth
temperatures or annealing) which leads to gaussian profiles.
Hence, real DMS heterostructures are more accurately said to
contain ``quasi-2D" distributions of spins, with a corresponding
magnetization and degree of spin clustering somewhere between that
of bulk (3D) and planar (2D) spin distributions.

The local, planar magnetic concentration in these quasi-2D spin
distributions varies significantly from monolayer to monolayer,
strongly affecting the probability of forming spin clusters (which
themselves may span many monolayers). It is desirable to
quantitatively predict the degree of this spin clustering in a
given DMS heterostructure so that accurate comparisons can be made
with real data. In this paper we present exact expressions for
determining the number and type of spin clusters (singles, pairs,
open- and closed triples) for {\it arbitrary} distributions of
magnetic spins in the common (100) growth direction. The results
reveal a rather surprising insensitivity of the computed
magnetization to the form of the intermixing profile
(exponential/gaussian), and highlight important limits on the
maximum possible magnetization using MBE techniques. High-field
photoluminescence (PL) and reflectivity studies of DMS
superlattices and quantum wells containing quasi-2D magnetic
planes are compared with the analytic results.

Spin clustering in DMS derives predominantly from the strong
antiferromagnetic {\it d-d} exchange between nearest-neighbor (NN)
magnetic cations (${J_{NN}\sim-10K}$). As outlined in the work of
Shapira\cite{Shapira} and others\cite{Wang}, single Mn${^{2+}}$
cations with no magnetic NNs are ${S=\frac{5}{2}}$ paramagnets,
with Brillouin-like magnetization. Two NN Mn${^{2+}}$ cations form
an antiferromagnetically-locked pair with zero spin at low
magnetic fields, and step-like magnetization at high fields and
low temperatures. Three Mn${^{2+}}$ spins can form a closed or
open triple with net spin ${S_T=\frac{1}{2}}$  and
${S_T=\frac{5}{2}}$ (respectively) at low fields, and a unique set
of magnetization steps at high fields.\cite{Wang} Spins in higher
order clusters are usually treated empirically and often exhibit a
linear susceptibility at high magnetic fields.\cite{Shapira}

The magnetization of monolayer planes of Mn${^{2+}}$ spins is a
significant challenge for conventional magnetometry.
Alternatively, the magnetization from DMS heterostructures may be
inferred from their giant magneto-optical properties. The
${J_{sp-d}}$ exchange interaction between electrons/holes and
local Mn${^{2+}}$ moments generates giant exciton spin-splittings
that are proportional to the {\it local} magnetization within the
exciton wavefunction. Using the giant spin-splitting of confined
excitons to probe the magnetization within a quantum well, the
studies of Gaj\cite{Gaj}, Grieshaber\cite{Grieshaber}, and of
Ossau\cite{Ossau} clearly established i) an enhanced paramagnetic
response in very thin layers of magnetic semiconductor, and ii)
that ``ideal" magnetic-nonmagnetic semiconductor interfaces are
smeared out due to segregation of Mn${^{2+}}$ during growth. A
clear example of both effects can be seen in the high-field PL
data of Figure 1.
\begin{figure}
\epsfxsize=3.0in \center \epsfbox{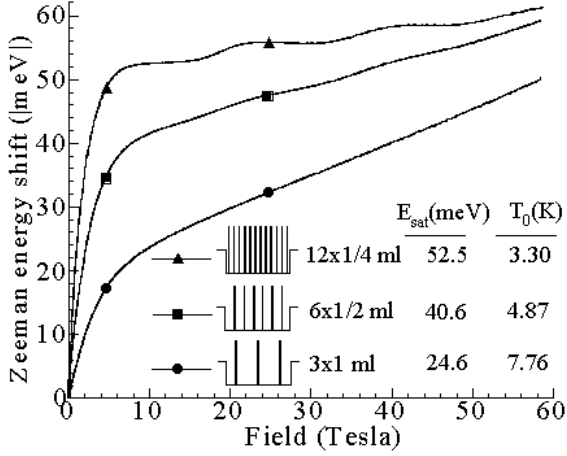} \vspace{0.1in}
\caption{Measured energy shift (abs. value) of the exciton PL to
60 T at 350mK in three quantum wells containing the {\it same
number} of Mn${^{2+}}$ spins, but in different quasi-2D
distributions, showing increased spin clustering with increasing
2D concentration.} \label{fig1}
\end{figure}
Here, we measure the giant energy shift ($\propto$ magnetization)
of the band-edge exciton PL to 60 Tesla in three quantum wells,
each containing the {\it same total number} of Mn${^{2+}}$ spins,
but in very different {\it distributions}. The samples are
120$\AA$ ZnSe/Zn${_{.8}}$Cd${_{.2}}$Se single quantum wells into
which the magnetic semiconductor MnSe has been incorporated in the
form of `digital' planes of 25\%, 50\%, and 100\% monolayer
coverage (12, 6, and 3 planes, respectively). The samples and the
experimental method have been described elsewhere.\cite{Crooker}
Evidence of decreased spin clustering with decreasing planar
concentration is clear in the low-field magnetization (${H<8T}$),
which is largely due to isolated Mn${^{2+}}$ spins and is fit to a
modified Brillouin function,
${E_{sat}=B_{5/2}[5\mu_BH/k_B(T+T_0)]}$. ${E_{sat}}$ is the
saturation splitting and ${T_0}$ is an empirical parameter that
accounts for long-range Mn-Mn interactions.\cite{Furdyna} As
shown, as the planar Mn${^{2+}}$ density increases, ${E_{sat}}$
decreases while ${T_0}$ grows, consistent with the expectation
that increasing the planar density results in fewer single
(un-clustered) Mn${^{2+}}$ spins and more long-range correlations
between Mn${^{2+}}$ spins. Further, with increasing planar
density, the high-field susceptibility evolves from magnetization
steps (from Mn-Mn pairs), to the linear susceptibility common in
large, highly correlated spin clusters. The presence of low-field
paramagnetism in the 100\% MnSe planes suggest, however, the
quasi-2D nature of these spin distributions: a full (perfect) 2D
magnetic monolayer would behave as one infinite cluster and
contain no paramagnetic spins whatsoever.
\begin{figure}
\epsfxsize=3.1in \center \epsfbox{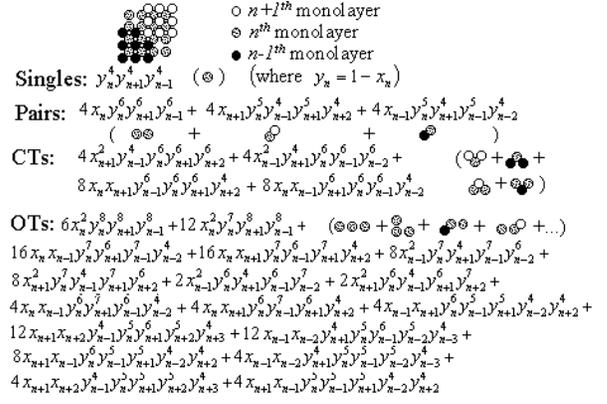} \vspace{0.1in}
\caption{Probability of clustering for a spin in the ${n^{th}}$
(100) monolayer of an arbitrary Mn${^{2+}}$ distribution.
Multiplying by ${x_n}$ gives the {\it number} of clustered spins.
Cations in the ${n+1^{th}}$, ${n^{th}}$, and ${n-1^{th}}$
monolayer are labeled by white, shaded, and black dots
respectively.} \label{fig2}
\end{figure}

The probability of forming a particular spin cluster is
essentially determined by the number of (possibly magnetic) NNs
bordering the cluster, and the number of ways the cluster can
form. In the zincblende crystal structure we consider, cations
form an fcc lattice, with twelve NNs per cation.  Thus the
probability of a Mn${^{2+}}$ being single or paired in bulk
crystals (3D) is ${P^{3D}_s=(1-x)^{12}}$  and
${P^{3D}_p=12x(1-x)^{18}}$ respectively, where ${x}$ is the
Mn${^{2+}}$ concentration. For perfect 2D monolayers grown in the
(100) direction, the cations form a 2D square lattice with only
four possible magnetic NNs per cation, so that
${P^{2D}_s=(1-x)^4}$ and ${P^{2D}_p=4x(1-x)^6}$. In a real system,
the effects of diffusion and/or segregation intermix adjacent
monolayers, so that a ``perfect" 2D plane of DMS is smeared over
several monolayers, with the ${n^{th}}$ (100) monolayer having a
2D magnetic concentration ${x_n}$ (assumed to be random within the
plane). Clustering within these quasi-2D spin distributions can be
modeled numerically\cite{Harrison,Ossau} or through analytic
approximations\cite{Grieshaber}, but an exact expression has been
lacking. Figure 2 shows the exact probabilities of a Mn${^{2+}}$
spin in the  ${n^{th}}$ monolayer belonging to a single, pair,
closed- and open triple. Diagrams show the clusters under
consideration - {\it e.g.}, there are three different types of
Mn${^{2+}}$ pairs (with the paired spin in the ${n-1^{th}}$,
${n^{th}}$, or ${n+1^{th}}$ monolayer), each four-fold degenerate.
There are four types of closed triples (for a total of 24), and
126 total configurations for open triples. (We do not attempt the
1900 configurations of spin quartets that have been recently
identified in the bulk\cite{Liu}, nor do we consider the much
weaker distant-neighbor couplings between Mn${^{2+}}$
moments.\cite{Bindilatti})

This algorithm allows for an exact calculation of spin clusters in
a heterostructure with an {\it arbitrary} distribution of
Mn${^{2+}}$ in the (100) direction. An example of its utility is
shown in Fig. 3, where we compute the number of Mn${^{2+}}$ in
singles, pairs, triples, and higher-order clusters in a 10
monolayer (10 ml) wide nonmagnetic quantum well with magnetic
(${x_{Mn}=30\%}$) barriers. We assume full segregation of atoms
during growth, giving ${e^{-n/\lambda}}$ and ${1-e^{-n/\lambda}}$
Mn${^{2+}}$ profiles (${\lambda\sim1.44 ml}$) at the first and
second interfaces (Fig 3a). Figs. 3b-e show the type and number of
clusters in each monolayer. Although the Mn${^{2+}}$ density is
comparatively small near the center of the quantum well, the
paramagnetic contribution from single spins to the {\it
optically-measured} magnetization can be significant, as it
depends on their overlap with the exciton wavefunction as shown.
The density of triples and of pairs is clearly peaked in the
quasi-2D interfaces. In the barriers, the spin distribution is
bulk-like and the vast majority of spins are bound up in higher
order clusters that contribute little to the low-field
magnetization.\begin{figure} \epsfxsize=3.0in \center
\epsfbox{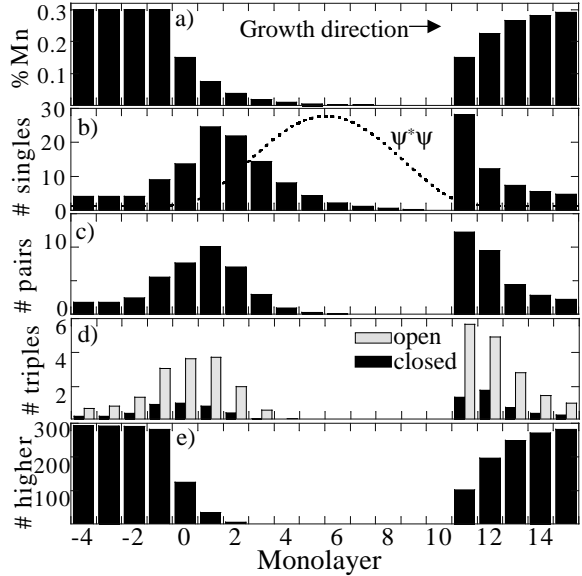} \vspace{0.1in}\caption{Schematic of the
Mn${^{2+}}$ concentration, including segregation effects, in a 10
monolayer wide quantum well with ${x_{Mn}}$=30\% magnetic
barriers.  b-e) The calculated number of Mn${^{2+}}$ cations (per
thousand sites) in singles, pairs, open/closed triples, and higher
order spin clusters.} \label{fig3}
\end{figure}

Modeling quasi-2D spin distributions leads to some rather
unexpected results.  In particular, it is clear that magnetization
studies alone will be of limited use in distinguishing the exact
{\it form} of the intermixing profile. Fig. 4a shows the
calculated magnetization for four different profiles of an
initially 2D monolayer containing 20\% Mn${^{2+}}$, where we
include the magnetization from singles, pairs, triples, and higher
order clusters following Refs. 1 and 6. Though unrealistic, the
first two profiles - a perfect 2D plane with ${x_{Mn}=20\%}$ and
two adjacent planes with ${x_{Mn}=10\%}$ -- illustrate an
important point: clustering often "conspires" to equalize
low-field magnetizations. Although the single monolayer contains
5\% fewer single Mn${^{2+}}$ spins, it contains over a third {\it
more} open triples and higher-order clusters, which act to
equalize the deficit.
\begin{figure} \epsfxsize=3.0in \center
\epsfbox{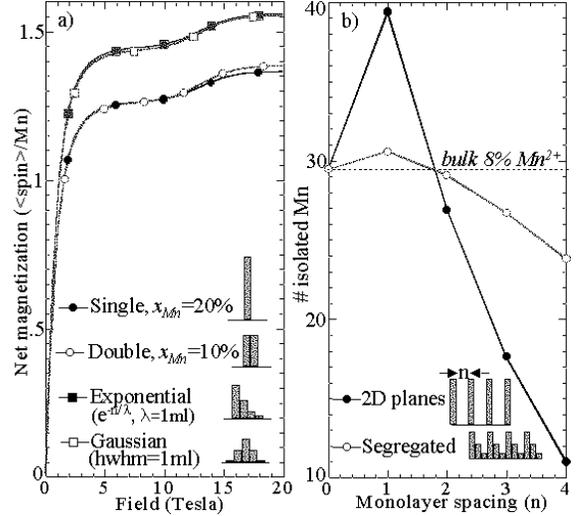} \vspace{0.1in} \caption{a) Calculated
average magnetization (per spin) for the Mn${^{2+}}$ spin profiles
shown.  Despite differing intermixing profiles, the magnetization
is often indistinguishable. b) Calculated number of isolated
Mn${^{2+}}$ spins (per thousand sites) assuming a 8\% bulk spin
distribution is redistributed as digital planes (see text). Only
if planes are spaced every other monolayer is there an
enhancement, although segregation greatly reduces the effect.}
\label{fig4}
\end{figure}
Only at the first magnetization step are the profiles
distinguishable, as the single monolayer contains fewer Mn-Mn
pairs. The last two profiles represent the exponential and
gaussian profiles roughly expected from segregation and diffusion,
respectively, with decay length and half-width equal to 1 ml.
Again, the calculated magnetizations are nearly identical
(although larger than for the first two profiles). Thus,
magnetization measurements alone cannot distinguish the form of
the spin profile. However, {\it assuming} a particular form, the
magnetization does depend sensitively on the segregation (or
diffusion) length, which can then be used to fit an intermixing
lengthscale as demonstrated below.

The model we present can also identify configurations for
realizing the maximum possible magnetization per unit volume in
MBE-grown structures. One motivation for growing `digital' alloys
is to exploit the reduced clustering of 2D planes to achieve
enhanced magnetizations beyond those possible with bulk, 3D
distributions.\cite{Crooker,Fiederling} In bulk DMS, the maximum
paramagnetic response is obtained with ${x_{Mn}\sim8\%}$, where
isolated Mn${^{2+}}$ spins comprise ${\sim2.9\%}$ of all cation
sites. In Fig. 4b we investigate whether it is then possible --
{\it with the same total number of Mn${^{2+}}$ spins} -- to
increase the number of isolated spins by redistributing the
Mn${^{2+}}$ in digital planes (solid dots). Bulk can be thought of
as 2D planes of spins with ${x_{Mn}^{2D}}$=8\%, spaced every
monolayer. Next, we consider 2D planes with twice the density
(${x_{Mn}^{2D}}$ =16\%), spaced every other monolayer, which
results in a paramagnetic enhancement of over a third, as shown.
However, spacing planes with ${x_{Mn}^{2D}}$=24\% every third
monolayer results in {\it fewer} free Mn${^{2+}}$ spins per unit
volume than in the case of bulk. Additional divisions continue to
reduce the paramagnetic response. So, only by spacing magnetic
planes every other monolayer is it possible to increase the
density of free Mn${^{2+}}$ beyond 3D spin distributions. However,
{\it any} intermixing during growth couples the 2D planes and
dramatically reduces the paramagnetic enhancement, as shown by the
open dots for the case of full segregation. Of course, clever
schemes for control of the spin distribution {\it within} the 2D
plane could certainly result in reduced spin clustering, such as
MBE growth in the (120) direction, where neighboring cation sites
in the (120) plane are {\it not} nearest
neighbors.\cite{Fiederling} Thusfar, however, such efforts have
been hampered by the inevitable inter-monolayer mixing of atoms
during growth, leading to spin clusters.

We apply the model to measurements of superlattices and quantum
wells containing ``digital" planes of DMS. Fig. 5a shows the
measured splitting between exciton spin states in two
superlattices with nominally single monolayers of
Zn${_{.75}}$Mn${_{.25}}$Se and Zn${_{.50}}$Mn${_{.50}}$Se
(separated by 4ml of ZnSe). The dotted lines are Brillouin fits to
the low-field magnetization (${H<8T}$). Increased spin clustering
in the Zn${_{.50}}$Mn${_{.50}}$Se monolayers is evident in the
smaller paramagnetic saturation, and more linear high-field
susceptibility. With perfect 2D planes, however, it is impossible
to account for the 15\% larger paramagnetic saturation from the
superlattice with Zn${_{.75}}$Mn${_{.25}}$Se planes. However,
assuming exponential, segregated Mn${^{2+}}$ profiles ${\propto
e^{-n/\lambda}}$ for each of the Zn${_{1-x}}$Mn${_x}$Se planes
(reasonable for the low growth temperature of 300 C), the relative
low field saturations can be reproduced with a decay length
$\lambda$=1.15ml, implying partial segregation during growth. As a
final study (Fig. 5b), we attempt to account for the size of the
magnetization steps observed in PL from the quantum well
containing twelve 1/4ml planes of MnSe. Magnetization steps arise
from the partial unlocking of antiferromagnetically-bound Mn-Mn
pairs, resulting in a step height proportional to the number of
pairs. The observed magnetization steps are {\it never more than}
5\% of the low-field `saturation magnetization' ${M_{sat}}$, a
ratio which is much smaller than predicted by {\it any}
conceivable distribution profile of the Mn${^{2+}}$ within the
quantum well. The expected step height for 3D, 2D, and segregated
2D spin distributions are shown for comparison. This puzzling
anomaly is seen in all `digital' samples, and even quantum wells
containing bulk (${x_{Mn}^{3D}}$ =8\%) DMS show a similar deficit.
We postulate this effect is due to the nature of the PL
measurement itself, which is not a direct measure of
magnetization, but is rather only proportional to the
magnetization through the ${J_{sp-d}}$ exchange interaction and
the Mn${^{2+}}$-exciton wavefunction overlap. It is anticipated
that true magnetization studies will reveal the correct magnitude
of the magnetization step.
\begin{figure}
\epsfxsize=3.0in \center \epsfbox{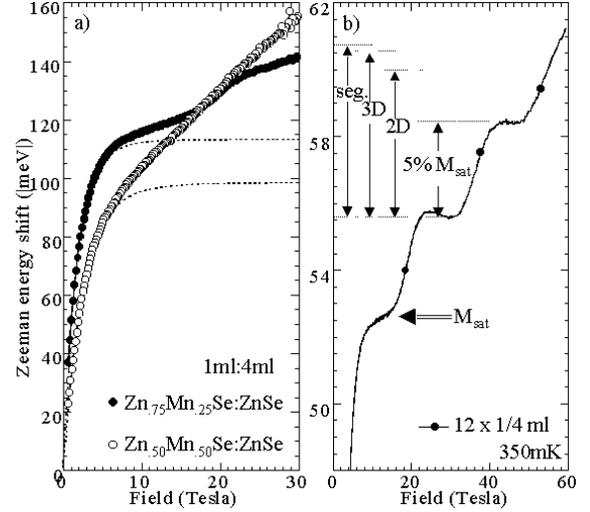} \vspace{0.1in}
\caption{a) Measured spin splitting at 1.6K for the two
superlattices shown, from reflectivity data.  Dotted lines are
Brillouin function fits to the low-field data.  b) The measured
magnetization steps (via PL) in the quantum well with 1/4 ml
magnetic planes.  The steps are smaller than any Mn${^{2+}}$
distribution would predict.} \label{fig5}
\end{figure}

In summary, we have presented a method for calculating the exact
number of spin singles, pairs, triples, and higher order clusters
for an arbitrary magnetic concentration profile in the (100)
growth direction, to model the magnetic properties of real,
quasi-2D spin distributions in DMS heterostructures. Calculation
of the magnetization for diffusion and segregation profiles
reveals nearly identical values, so that fitting an intermixing
length is likely possible only when the form (exponential,
gaussian, etc) of the quasi-2D profile is assumed {\it a priori},
as was demonstrated for the case of ZnMnSe:ZnSe superlattices. The
model also predicts a larger paramagnetism compared with bulk spin
distributions {\it only} if digital planes are spaced every other
monolayer, although the effects of intermixing will greatly reduce
any enhancement. Lastly, the discrepancy between the magnitude of
observed and predicted magnetization steps remains outstanding.
The methods outlined in this paper will be of use in modeling
future epitaxially-grown DMS heterostructures, where spin
distributions can be engineered with nearly monolayer precision.

The authors gratefully acknowledge the assistance of J. Schillig
and M. Gordon during operation of the 60T Long-Pulse magnet.  Work
supported by grants NSF DMR 97-01072 and 9701484.

\end{document}